\documentclass[preprint,prb,aps]{revtex4}
\usepackage{amsmath,amssymb}
\usepackage{graphicx}
\usepackage{dcolumn}
\usepackage{bm}

\begin{document}
\title[Position dependent mass]{Quantum solvability of a general ordered position dependent mass system: Mathews-Lakshmanan oscillator}

\author{S. Karthiga$^{\dagger}$, V. Chithiika Ruby$^{\ddagger}$, M. Senthilvelan$^{\dagger}$ and M. Lakshmanan$^{\dagger}$}

\affiliation{$^{\dagger}$ Centre for Nonlinear Dynamics, Bharathidasan University,  Tiruchirappalli 620 024, Tamil Nadu, India}
\address{$^{\ddagger}$ Department of Physics, Cauvery College for Women, Tiruchirappalli 620 018, Tamil Nadu, India.}

\begin{abstract}
In position dependent mass (PDM) problems, the quantum dynamics of the associated systems have been understood well in the literature for particular orderings.  However, no efforts seem to have been made to solve such PDM problems for general orderings to obtain a global picture.  In this connection, we here consider the general ordered quantum Hamiltonian of an interesting position dependent mass problem, namely the Mathews-Lakshmanan oscillator, and try to solve the quantum problem for all possible orderings including Hermitian and non-Hermitian ones. The other interesting point in our study is that for all possible orderings, although the Schr\"odinger equation of this Mathews-Lakshmanan oscillator is uniquely reduced to the associated Legendre differential equation, their eigenfunctions cannot be represented in terms of the associated Legendre polynomials with integral degree and order.  Rather the eigenfunctions are represented in terms of associated Legendre polynomials with non-integral degree and order. We here explore such  polynomials  and represent the discrete and continuum states of the system.  We also exploit the connection between associated Legendre polynomials with non-integral degree with other orthogonal polynomials such as Jacobi and Gegenbauer polynomials.  
\end{abstract}

\maketitle

\section{Introduction}
\par Studies on the quantum dynamics of position dependent mass Schr\"odinger equation have attracted wide interest over the years \cite{ml,pdd1,pdd2,pdd3,sen1,sen2}.  The reason is that this type of effective mass Schr\"odinger equations are helpful in studying optical and electronic properties of semiconductors \cite{semi}, quantum dots \cite{qd1, qd2}, quantum wells \cite{qd1,qd2, qd3, qw1}, quantum liquids \cite{ql1} and super-lattice band structures \cite{sup}.  The important problem that one faces while considering such effective mass Schr\"odinger equation is the problem of ordering of the kinetic energy operator \cite{ruby}.  The momentum and mass functions associated with the kinetic energy operator in such position dependent mass problems do not commute with each other.   Consequently the kinetic energy operator in these cases can be written or ordered in multiple ways.  This ordering ambiguity is a long standing problem in quantum physics. 
\par In the literature, many ordering schemes have been proposed and studied.  For example one can cite the orderings proposed by Ben Daniel and Duke \cite{bdd}, Gora and Williams \cite{gora}, Zhu and Kroemer \cite{zhu}, Morrow and Brownstein \cite{mor}, Li-Kuhn \cite{li}, Weyl \cite{weyl}, von-Roos \cite{von} and so on.  In the above types of orderings, the Hamiltonians were restricted to be Hermitian, whereas recent studies on quantum systems allow the possibility that the Hamiltonians can be even non-Hermitian.  Complete real energy spectra have been observed in the cases of such non-Hermitian Hamiltonians so that one can also consider the orderings that are non-Hermitian too. Taking into account the above fact, a more general ordered Hamiltonian has been proposed in \cite{trab} and the work shows that infinite number of orderings are possible for all PDM systems. In the literature, one can find that the quantum dynamics of PDM problems are mostly studied by considering a particular ordering.  No efforts seem to have been taken to solve a PDM problem for all possible orderings. 
\par In this article, we consider an interesting PDM system, namely the Mathews-Lakshmanan oscillator, and try to solve it for all possible orderings.  The Mathews-Lakshmanan (ML) oscillator is a non-polynomial oscillator and has attracted considerable attention over the years \cite{classi,cl4,cl2,cl5,cl6,cl7,cl8,cl9,cl10,cl11,sen3} from different perspectives.  The quantum solvability of this model has also been studied for particular orderings \cite{ml, cari}.  In these studies, it has also been shown that the Schr\"odinger equation of the system can be reduced to the associated Legendre differential equation or to a $\lambda$-dependent Hermite differential equation (where $\lambda$ is one of the parameters of the system) and that the Mathews-Lakshmanan oscillator admits a nonlinear energy spectrum.  However, the above Mathews-Lakshmanan oscillator has not been solved for all possible orderings and the quantum dynamics corresponding to different orderings is still an open problem.
\par The other interesting point one can note is that the Mathews-Lakshmanan oscillator can be reduced to the associated Legendre differential equation for all possible orderings.  But, the eigenfunctions of the system cannot be always written in terms of the associated Legendre polynomials with integral degree and order ($P_{\nu}^{\mu}$, the degree $\nu=1,2,3...$ and the order $\mu=1,2,3...$); rather it is represented in terms of associated Legendre polynomials with non-integral degree and order ($P_{\nu}^{\mu}$, $\nu \in R$, $\mu \in R$) or fractional order associated Legendre polynomials.   We here explore these polynomials and present their interconnection with the $\lambda$-dependent Hermite polynomials and Jacobi polynomials.  The existence of interesting continuum bound states is also pointed out. 
\par  To demonstrate the above facts, the manuscript is structured in the following way.  In Sec. \ref{genn}, we present the form of the general ordered Hamiltonian. In Sec. \ref{mlo_lit}, we give an introduction to Mathews-Lakshmanan oscillator and review the works that have been carried out on the quantum dynamics of this oscillator.  In Sec. \ref{lamg0}, we discuss the quantum solvability of this oscillator for the case $\lambda>0$, where $\lambda$ is the system parameter.  We demonstrate the use of associated Legendre polynomials with non-integral degree and order in expressing the eigenfunctions of the system and show the orthonormality of the eigenfunctions.  Similarly in Sec. \ref{laml0}, we have detailed the quantum solvability of the system in the case $\lambda<0$.  Finally, we summarize our results in Sec. \ref{sumry}.
 
\section{\label{genn} General ordered Hamiltonian}
\par The non-commutativity of momentum operator with position dependent mass function in PDM problems allows the kinetic energy operator to be written in multiple ways.  The literature evidences the consideration of various ordering schemes, namely Ben-Daniel and Duke ordering \cite{bdd}, Gora and Williams ordering \cite{gora}, Zhu and Kromer's ordering \cite{zhu}, Weyl ordering \cite{weyl} and so on.  In a similar manner, there have been efforts to find the general ordered kinetic energy operator.  Importantly, von-Roos first proposed a two-parameter general ordering \cite{von}, where the operator $\hat{T}$ is expressed as
\begin{eqnarray}
\hat{T}=\frac{1}{4}\,\left[m^\alpha\,\hat{\bf p}\,m^\beta\,\hat{\bf p}\,m^\gamma + m^\gamma\,\hat{\bf p}\,m^\beta\,\hat{\bf p}\,m^\alpha \,\right],  \label{H}
\end{eqnarray}
where $\alpha,$ $\beta$ and $\gamma$ are arbitrary ordering parameters restricted by $\alpha+\beta+\gamma=-1$.  The above form of ordering has been found to include all the orderings mentioned earlier except for the Weyl ordering.  By including Weyl ordering, the above von-Roos ordering has been extended to the following form \cite{von2}
\begin{eqnarray}
\hat{T}=\frac{1}{4(a+1)}\left[a\left(\frac{1}{m}\,\hat{\bf p}^2 + \hat{\bf p}^2\,\frac{1}{m}\right)+m^\alpha \hat{\bf p}\,m^\beta \hat{\bf p}\,m^\gamma + m^\gamma \hat{\bf p}\,m^\beta \hat{\bf p}\,m^\gamma\,\right],\label{I}
\end{eqnarray}
where $a$ is an arbitrary parameter \cite{von2}.  Although Eq. (\ref{I}) has been considered as a more general ordered form of kinetic energy operator, it is still not sufficiently general.  There are two reasons to state the above.  Firstly, one can note that in Eqs. (\ref{H}) and (\ref{I}) the terms have been written as a combinations of two or four ordered terms.  But realistically, we can have combinations of a large number of terms with different orderings.  Thus the kinetic energy operator in (\ref{I}) cannot be considered to be the most general one. 
\par The second reason is that the above mentioned orderings are Hermitian orderings, where they preserve the Hamiltonian to be Hermitian (${\cal{H}}^\dagger={\cal{H}}$).   Due to the belief that only Hermitian Hamiltonians can give rise to real energy spectra, such focus on Hermitian orderings did exist.  The recent studies on non-Hermitian Hamiltonians reveal that there also exist non-Hermitian Hamiltonians which support real spectra \cite{bend}.   Thus the position dependent mass systems can also be studied with non-Hermitian orderings.  Due to the above mentioned two reasons, the general ordered form of kinetic energy operator can be written as
\begin{eqnarray}
\hat{T}=\frac{1}{2}\sum_{i=1}^{N}w_i m^{\alpha_i} \hat{\bf p}\,m^{\beta_i} \hat{\bf p}\,m^{\gamma_i},\label{J}
\end{eqnarray}
where $N$ is an arbitrary integer, $w_i$ is the weight parameter satisfying the condition ${\sum}_{i=1}^{N} w_i=1$.  Also, $\alpha_i$, $\beta_i$ and $\gamma_i$ are ordering parameters constrained by $\alpha_i+\beta_i+\gamma_i=-1$.  It is interesting to note that the above kinetic energy operator can also be simplified as 
\begin{eqnarray}
\hat{T}=\frac{1}{2} \hat{\bf p}\,\frac{1}{m}\,\hat{\bf p}+({\overline\gamma}-{\overline\alpha}) \frac{i\hbar}{2}\, \,\left(\vec{\bf \nabla}\frac{1}{m}\right).\hat{\bf p} + \frac{\hbar^2}{2}\left[\overline\gamma\,\nabla^2 \left(\frac{1}{m}\right)+\overline{\alpha\gamma}\,\left(\vec{\bf \nabla}  \frac{1}{m}\,\right)^2\ m\right], \quad\label{K}
\end{eqnarray}
where 
\begin{eqnarray}
\bar{\alpha}=\sum_{i=1}^{N}w_i \alpha_i, \;\; \bar{\gamma}=\sum_{i=1}^{N}w_i \gamma_i , \;\; \overline{\alpha\gamma}=\sum_{i=1}^{N}w_i\alpha_i\gamma_i.
\label{alb}
\end{eqnarray}
denote the weighted mean values. The dependence of $\bar{\beta}$ is removed through the constraint $\bar{\alpha}+\bar{\beta}+\bar{\gamma}=-1$.  From Eq. (\ref{K}), the Hamiltonian can be written in the simpler form as
\begin{eqnarray}
\hat{H}=\frac{1}{2}\,\hat{ \bf p}\,\frac{1}{m}\,\hat{\bf p}+(\overline\gamma-\overline\alpha)\frac{i\hbar}{2}\,\vec{\nabla} \left(\frac{1}{m}\,\right).\hat{\bf p}+V_{eff},
\label{hveff}
\end{eqnarray} 
where $V_{eff}$ is the effective potential that is of the form
\begin{eqnarray}
V_{eff}=\frac{\hbar^2}{2}\,\left[\overline\gamma\,{\bf \nabla}^2 \left(\frac{1}{m}\right)+\overline{\alpha\gamma}\,\left(\vec{\bf \nabla} \frac{1}{m}\,\right)^2 m \right] +V.
\label{veff}
\end{eqnarray} 
  It is now obvious that the kinetic energy operator considered in Eqs. (\ref{J}) or (\ref{hveff}) is really more general and it includes all possible ordering schemes. Note that for $\bar{\alpha}=\bar{\gamma}$, the ordering given in Eq. (\ref{hveff}) becomes Hermitian and for $\bar{\alpha} \neq \bar{\gamma}$, it is non-Hermitian.  Eqs. (\ref{J}) and (\ref{hveff}) clearly show that the Schr\"odinger equation corresponding to the different orderings are different and, consequently, the energy eigenvalues and eigenfunctions corresponding to these cases are also different.  Thus it is of great interest to look upon how the energy spectrum and eigenfunctions vary with respect to different orderings. For this purpose, we here consider a specific position dependent mass system and try to exactly solve the system for the general ordering given in (\ref{hveff}).  
\section{\label{mlo_lit}Mathews-Lakshmanan oscillator}
We here consider a remarkable position dependent mass problem, namely the Mathews-Lakshmanan oscillator, whose classical Hamiltonian can be written as
\begin{eqnarray}
{\cal{H}}=\frac{p^2}{2 m(x)}+V(x),
\label{hamil}
\end{eqnarray}
where the spatially varying mass and potential take the forms
\begin{eqnarray}
m(x)=\frac{1}{1-\lambda x^2}, \quad V(x)=\frac{k x^2}{2(1-\lambda x^2)}.
\label{mxv}
\end{eqnarray}
Note that in the above, we assume $m=1$ when $\lambda=0$, for convenience, which is followed throughout the present paper.  The above Mathews-Lakshmanan oscillator may be considered as the zero-dimensional version of a scalar nonpolynomial field equation \cite{classi,ml}.   The literature shows evidence for a large interest over the system both from classical and quantum points of views \cite{classi,cl2,cl4,cl5,cl6,cl7,cl8,cl9,cl10,cl11}.  The above oscillator is often considered as a nonlinear extension of the harmonic oscillator as the Hamiltonian in (\ref{hamil}) tends to the harmonic oscillator Hamiltonian in the limit $\lambda \rightarrow 0$.  The other interesting aspect of this nonpolynomial nonlinear oscillator is that classically this system exhibits simple harmonic oscillations with amplitude dependent frequency \cite{classi} where the exact solution of the system can be simply written as 
\begin{eqnarray}
x(t)=A \sin(\omega t+\delta), \quad \omega=\sqrt{\frac{\omega_0^2}{1+\lambda A^2}}. 
\end{eqnarray}  The Schr\"odinger equation corresponding to the system has been solved for particular orderings \cite{ml,cari}, where
\begin{enumerate}
\item the original effort has been taken by Mathews and Lakshmanan themselves and the ordering considered by them is a Hermitian ordering \cite{ml} where
\begin{eqnarray}
\hat{T}=\frac{1}{2}\left(\hat{p}^2 \frac{1}{2m(x)}+\frac{1}{2m(x)} {\hat{p}^2}\right).
\label{ml_or}
\end{eqnarray}
In other words, the ordering can be defined by  $\alpha_1=0$, $\gamma_1=-1$, $\alpha_2=-1$, $\gamma_2=0$ and the ordering weights in Eq. (\ref{J}) are $w_1=w_2=\frac{1}{2}$.  Substituting these in Eq.(\ref{alb}), we can obtain $\bar{\alpha}=-1/2$,  $\bar{\gamma}=-1/2$ and $\overline{\alpha \gamma}=0$.  Note that $\bar{\alpha}=\bar{\gamma}$, defines the ordering as Hermitian. 
For this ordering, Mathews and Lakshmanan obtained the solutions of the time independent Schr\"odinger equation in terms of associated Legendre polynomials with non-integral degree and order.  
\item Later, Carinena {\it et al} considered a non-Hermitian ordering of the form \cite{cari}
\begin{eqnarray}
\hat{T}=\frac{1}{2 } \frac{1}{\sqrt{m(x)}}\hat{p}\frac{1}{\sqrt{m(x)}}\hat{p},
\label{car_or}
\end{eqnarray}
where the ordering can be seen as $\alpha_1=\bar{\alpha}=-1/2$ and $\gamma_1=\bar{\gamma}=0$ in Eq. (\ref{J}). 
 Due to the consideration of Mathews-Lakshmanan oscillator as a nonlinear extension of harmonic oscillator, Carinena {\it et al} were interested in expressing the eigenfunctions in terms of $\lambda$-dependent Hermite polynomials.
\end{enumerate} 
\par Now, it is natural to ask how would the eigenfunctions and eigenvalues differ from one another in different orderings when the general ordering problem is considered?  Will the problem be exactly solvable for all orderings?  Whether the eigenfunctions corresponding to different orderings can be represented uniquely by an orthogonal polynomial or different orthogonal polynomials need to be used to represent different orderings?  Will the polynomials used by Mathews and Lakshmanan (associated Legendre polynomials with non-integral degree and order) and the one used by Carinena {\it et al} ($\lambda$-dependent Hermite polynomials) have any interconnections?  With the above questions in mind, we here discuss the solvability of the Mathews-Lakshmanan oscillator for all orderings.  Particularly, we detail here the interesting associated Legendre polynomials with non-integral degree and order and find the interconnection to other polynomials such as Jacobi and $\lambda$-dependent Hermite polynomials.
\par In addition to the above, the Mathews-Lakshmanan oscillator has been studied under two different situations, that is with $\lambda>0$ and $\lambda<0$.  Considering the case $\lambda>0$, one can find from (\ref{mxv}) that the system has singularities at $x =\pm \frac{1}{\sqrt{\lambda}}$ and $\pm \infty$. Thus we can consider three regions, region-$1$: $-\infty \le x \le -\frac{1}{\sqrt{\lambda}}$, region-$2$: $-\frac{1}{\sqrt{\lambda}} \leq x \leq  \frac{1}{\sqrt{\lambda}}$ and region-$3$: ${\frac{1}{\sqrt{\lambda}}} \le x \le \infty$.  Due to the above, we can consider two situations.  Firstly, the system may be bounded in the region-$2$, where
\begin{eqnarray}
&\psi(x)=0& \quad \qquad \mathrm{for} \quad -\infty \le x \le -\frac{1}{\sqrt{\lambda}}, \;\; \mathrm{and} \;\; \frac{1}{\sqrt{\lambda}} \le x \le \infty,  \nonumber \\
&\psi(x) \neq 0& \quad \qquad \mathrm{for} \quad -\frac{1}{\sqrt{\lambda}} < x < \frac{1}{\sqrt{\lambda}}.
\label{boun1}
\end{eqnarray} 
Secondly, we can consider the situation in which
\begin{eqnarray}
&\psi(x)= 0& \;\,  \mathrm{for} \;\, -\frac{1}{\sqrt{\lambda}} \le x \le \frac{1}{\sqrt{\lambda}},  \qquad \nonumber \\
&\psi(x)\neq 0& \;\, \mathrm{for} \;\, -\infty < x <-\frac{1}{\sqrt{\lambda}} \quad \mathrm{or} \quad \frac{1}{\sqrt{\lambda}} < x < \infty \;\, \mathrm{and} \;\, \psi(\pm \infty)=0. \quad
\label{boun2}
\end{eqnarray}
\par In the case $\lambda<0$, the system has singularities only at $x =\pm \infty$. Thus, the eigenfunctions satisfy the boundary conditions as $\psi(\pm \infty)=0$ and that 
\begin{eqnarray}
\psi(x) \neq 0     \quad  \mathrm{for} \quad -\infty < x < \infty.
\label{case3}
\end{eqnarray}
In this article, we are essentially interested to solve the general ordered Hamiltonian for all the above cases. 
\section{\label{lamg0}Quantum solvability: case $\lambda>0$}
\par The Schr\"odinger equation corresponding to the Mathews-Lakshmanan oscillator for the most general ordering can be obtained from Eq. (\ref{hveff}) and Eq. (\ref{mxv}) and it is of the form
\begin{eqnarray}
\left[-\frac{\hbar^2}{2} (1-\lambda x^2)\frac{d^2}{dx^2}-\hbar^2 (\bar{\gamma}-\bar{\alpha}-1) \lambda x\frac{d}{dx}+V_{eff}\right] \psi(x)=E \psi(x), \label{scho_mlo}  
\end{eqnarray}
where 
\begin{eqnarray}
V_{eff}=\frac{\hbar^2}{2} \left[-2 \lambda \bar{\gamma}+\frac{4 \overline{\alpha \gamma} \lambda^2 x^2}{(1-\lambda x^2)}\right]+\frac{k x^2}{2(1-\lambda x^2)}. \qquad \qquad.
\label{scho_p}
\end{eqnarray}
One can clearly find that by choosing appropriate values for $\bar{\alpha}$ (or $\alpha_i$'s), $\bar{\gamma}$ (or $\gamma_i$'s) and ${\overline{\alpha \gamma}}$, we can get the Schr\"odinger equation for particular ordering scheme. For example, the choice $\alpha_1=\gamma_2=0$, $\alpha_2=\gamma_1=-1$ and $\bar{\alpha}=\frac{\alpha_1+\alpha_2}{2}=-1/2$ and $\bar{\gamma}=\frac{\gamma_1+\gamma_2}{2}=-1/2$ can produce the ordering considered in Eq. (\ref{ml_or}) originally by Mathews and Lakshmanan \cite{ml} and the choice $\alpha_1=\bar{\alpha}=-\frac{1}{2}$ and $\gamma_1=\bar{\gamma}=0$ can produce the non-Hermitian ordering (\ref{car_or}) that was considered by Carinena {\it et al} \cite{cari}. 
\par One can find that the Schr\"odinger equation in (\ref{scho_mlo}) can be simplified to
\begin{eqnarray}
(1-\lambda x^2)\frac{d^2 \psi}{dx^2}+2 a \lambda x \frac{d \psi}{dx}+\left(b+\frac{c x^2}{(1-\lambda x^2)}\right) \psi=0,
\label{sim_mlo}
\end{eqnarray}
where
\begin{eqnarray}
 a=\bar{\gamma}-\bar{\alpha}-1, \;\; b=2 \lambda \bar{\gamma}+\frac{2 E}{\hbar^2} \quad \mathrm{and} \quad c=-\left(4 \overline{\alpha \gamma} \lambda^2+\frac{k}{\hbar^2}\right).
\end{eqnarray}
\par  By introducing the following transformations in both the independent and dependent variables in Eq. (\ref{sim_mlo}),
\begin{eqnarray}
\psi=(1-z^2)^{(\bar{\gamma}-\bar{\alpha})/2} \phi(z), \quad z=\sqrt{\lambda} x,
\label{tr_din}
\end{eqnarray}
we can reduce it to the associated Legendre differential equation of the form
\begin{eqnarray}
(1-z^2) \frac{d^2\phi}{dz^2}-2 z \frac{d \phi}{dz}+\left(\nu(\nu+1)-\frac{\mu^2}{(1-z^2)}\right) \phi=0.
\label{legeq}
\end{eqnarray}
In the above, we have taken
\begin{eqnarray}
\nu(\nu+1)=\frac{2 E}{\hbar^2 \lambda}+(\bar{\gamma}+\bar{\alpha})+\mu^2, \qquad  \mu=\frac{  \tilde{d}}{\hbar \lambda}, 
\label{cons}
\end{eqnarray}
in which $\tilde{d}=\sqrt{k+\hbar^2 \lambda^2 \left((\bar{\gamma}-\bar{\alpha})^2+ 4 \overline{\alpha \gamma}\right)}$.  The above equation (\ref{cons}) indicates that the energy eigenvalues can be found easily from the expression for $\nu(\nu+1)$ as
\begin{eqnarray}
E=\frac{\hbar^2 \lambda}{2}\left(\nu(\nu+1) -\mu^2 -(\bar{\gamma}+\bar{\alpha})\right).
\label{engy}
\end{eqnarray}
Considering the eigenfunctions of the system, we recall the different situations mentioned in the previous section in Eqs. (\ref{boun1}) and (\ref{boun2}). In the following, we deduce the eigenfunction in the two cases, mentioned therein, separately. 
\subsection{Bound states in region-2}
\par First, we consider the situation in which the solution of the system exists only in the region-2 ( $-1 < z < 1$ or $-\frac{1}{\sqrt{\lambda}} < x < \frac{1}{\sqrt{\lambda}}$).  As the Schr\"odinger equation has been reduced to the associated Legendre differential equation, one may think that the eigenfunctions can be written in terms of the associated Legendre polynomials with integral order and degree as they form an orthonormal set inside [$-1,1$].  But the task is not that simple, as we cannot represent the eigenstates in terms of such associated Legendre polynomials. The reason is that in the latter case both $\nu$ and $\mu$ are found to be integers ($\nu=l=\mathrm{integer}$ and $\mu=m=\mathrm{integer}$).  But in our case, it is obvious from Eq. (\ref{cons}) that the parameter $\mu$ in general need not be an integer (Note that in the angular part of the Schr\"odinger equation of rotationally symmetric three dimensional problems, $\nu$ and $\mu$ can take only integer values).  Due to the above reason, we have to look for possible forms of $P_\nu^\mu$ or $Q_\nu^\mu$ or combinations of $P_\nu^\mu$ and $Q_\nu^\mu$ which remain finite everywhere inside $-1 \le z \le 1$. 
\par For the above purpose, let us first consider $P_\nu^\mu$ with $\mu >0$ (as both $P_\nu^\mu$ and $P_\nu^{-\mu}$ can be the solutions of associated Legendre differential equation, we consider only $\mu>0$), which can be expanded in terms of hypergeometric functions as \cite{book_ab}
\begin{small}
\begin{eqnarray}
P^{\mu}_{\nu}(z)= 2^{\mu} \sqrt{\pi} (1-z^2)^{\frac{-\mu}{2}}\left[ \left( \frac{F \left(-(\frac{\nu}{2} + \frac{\mu}{2}), \frac{1}{2} + \frac{\nu}{2} - \frac{\mu}{2};\frac{1}{2};z^2 \right)}{\Gamma{(\frac{1}{2} - \frac{\nu}{2} -\frac{\mu}{2})}\,\Gamma{(1 + \frac{\nu}{2} -\frac{\mu}{2}})}  \right) -\left(  \frac{2z F \left( \frac{1}{2} - \frac{\nu}{2} - \frac{\mu}{2},1 + \frac{\nu}{2} - \frac{\mu}{2};\frac{3}{2};z^2 \right)} {\Gamma{(\frac{1}{2} + \frac{\nu}{2} - \frac{\mu}{2})}\,\Gamma{(-\frac{\nu}{2} -\frac{\mu}{2}})} \right)\right],  \nonumber \\
   -1 <z<1. \quad  \label{hgf}
\end{eqnarray}
\end{small}
Note that in the above, $P_\nu^\mu(z)$ is not defined at $z=\pm 1$ (as it becomes singular at $z=\pm 1$) and the boundary conditions of the problem requires the eigenfunction to be zero at $z=\pm 1$.  In order to identify a well defined function at $z=\pm 1$, we replace $\mu$ by $-\mu$ as
\begin{small}
\begin{eqnarray}
 P^{-\mu}_{\nu}(z)= 2^{-\mu} \sqrt{\pi} (1-z^2)^{\frac{\mu}{2}}\left[ \left( \frac{F \left(-(\frac{\nu}{2} - \frac{\mu}{2}), \frac{1}{2} + \frac{\nu}{2} + \frac{\mu}{2};\frac{1}{2};z^2 \right)}{\Gamma{(\frac{1}{2} - \frac{\nu}{2} +\frac{\mu}{2})}\,\Gamma{(1 + \frac{\nu}{2} + \frac{\mu}{2}})}  \right) -\left(  \frac{2z F \left( \frac{1}{2} - \frac{\nu}{2} + \frac{\mu}{2},1 + \frac{\nu}{2} + \frac{\mu}{2};\frac{3}{2};z^2 \right)} {\Gamma{(\frac{1}{2} + \frac{\nu}{2} + \frac{\mu}{2})}\,\Gamma{(-\frac{\nu}{2} +\frac{\mu}{2}})} \right)\right],  \nonumber \\
  -1  \le z \le 1. \quad \label{hgf2}
\end{eqnarray}
\end{small}
As the associated Legendre differential equation is symmetric with respect to the change $\mu \rightarrow -\mu$, both $P_\nu^\mu$ and $P_\nu^{-\mu}$ are solutions of the associated Legendre differential equation.  By this replacement, we find $P_\nu^{-\mu}(z)$ becomes zero at $z=\pm 1$.
\par We further note here that $P_\nu^{-\mu}$ in Eq. (\ref{hgf2}) is a series solution.  But the bound states of a quantum problem are expressed by orthonormal set of eigenfunctions.  Due to this fact, we look for the situation in which $P_\nu^{-\mu}$ gives an orthonormal polynomial set.  For this purpose, we recall the polynomial condition on hypergeometric functions, where they can be expanded as
\begin{small}
\begin{eqnarray}
F(a,b;c;z) = 1 + \frac{ab}{c} z + \frac{a (a +1) b (b+1)}{c(c + 1)} \frac{z^2}{2!} +\frac{a (a +1)(a+2) b (b+1)(b+2)}{c(c + 1)(c+2)} \frac{z^3}{3!}+  ..... \quad \label{hf} 
\end{eqnarray}
\end{small}
The choice of $a=-n$ or $b=-n$, where $n=0,1,2,...,$ reduces the above series solution to an $n^{\mathrm{th}}$ order polynomial function.  Using this fact in the expression for $P_\nu^{-\mu}$ given by Eq.(\ref{hgf2}), we try to find the polynomial condition for it.  Whenever $-\frac{\nu}{2}+\frac{\mu}{2}=-n$ (or $\frac{1+\nu+\mu}{2}=-n$), the first term in Eq. (\ref{hgf2}) reduces to an $n^{\mathrm{th}}$ order polynomial function while the second term in Eq. (\ref{hgf2}) vanishes due to the term $\Gamma{(-\frac{\nu}{2}+\frac{\mu}{2})}$ (or $\Gamma{(\frac{1+\nu+\mu}{2})}$ ) in the denominator.  Similarly, when $ \frac{1}{2}-\frac{\nu}{2}+\frac{\mu}{2}=-n$ (or $1+\frac{\nu}{2}+\frac{\mu}{2}=-n$),  the second term in Eq. (\ref{hgf2}) gives rise to an $n^{\mathrm{th}}$ order polynomial function and the first term vanishes due to the term $\Gamma{(\frac{1}{2} - \frac{\nu}{2} +\frac{\mu}{2})}$ (or $\Gamma{(1+\frac{\nu}{2}+\frac{\mu}{2})}$) in the denominator.  Thus Eq. (\ref{hgf2}) can give polynomial solutions for  $-\frac{\nu}{2}+\frac{\mu}{2}=-n$  and $ \frac{1}{2}-\frac{\nu}{2}+\frac{\mu}{2}=-n$.  Combining these two criteria, we can write the polynomial condition for $P_\nu^{-\mu}$ as
\begin{eqnarray}
\nu=n+\mu,  \quad n=0,1,2,3,...
\label{poly}
\end{eqnarray} 
Thus, $P_{n+\mu}^{-\mu}$'s are found to be polynomials of order $n$ and they can be given by
\begin{eqnarray}
P_{n+\mu}^{-\mu}(z)&=&\frac{2^{-\mu} \sqrt{\pi} (1-z^2)^{\frac{\mu}{2}}}{\Gamma{(\frac{1}{2}-\frac{n}{2})}\Gamma{(1+\frac{n}{2}+\mu})} F\left(\frac{-n}{2},\frac{1}{2}+\frac{n}{2}+\mu; \frac{1}{2}; z^2\right), \quad n=0,2,4...
\label{even_n} \\
P_{n+\mu}^{-\mu}(z)&=&\frac{2^{-\mu}\sqrt{\pi}(1-z^2)^{\frac{\mu}{2}}(-2z)}{\Gamma{(\frac{1}{2}+\frac{n}{2}+\mu)}\Gamma{(\frac{-n}{2})}} F\left(\frac{1-n}{2}, 1+\frac{n}{2}+\mu; \frac{3}{2}; z^2\right), \quad n=1,3,5...
\label{odd_n}
\end{eqnarray}
\par One can also note from (\ref{hgf2}) that $P_{\nu}^{-\mu}$ can also give polynomial solutions for the conditions $\frac{1+\nu+\mu}{2}=-n$ and $1+\frac{\nu}{2}+\frac{\mu}{2}=-n$.  Combining these two criteria, one can get the polynomial condition as $\nu=-n-\mu-1$.  But as the associated Legendre differential equation is symmetric under the transformation $\nu \rightarrow -(\nu+1)$, the relation
\begin{eqnarray}
P_{-n-\mu-1}^{-\mu}(z)=P_{n+\mu}^{-\mu}(z)
\label{gg}
\end{eqnarray}
holds good.  
This condition implies that no new linearly independent solutions can be identified. 
\par  We can also look for possible linearly independent solutions from $Q_\nu^\mu$ but again find that there exists no other linearly independent solution.  Because, if one considers $Q_\nu^\mu$, we have \cite{book_ab}
\begin{eqnarray}
Q_\nu^\mu(z)=2^{\mu} \sqrt{\pi} (1-z^2)^{\frac{-\mu}{2}}\bigg[\frac{\cos{\frac{(\nu+\mu) \pi}{2}} \Gamma{(\frac{\nu}{2}+\frac{\mu}{2}+1)}z}{\Gamma{(\frac{\nu}{2}-\frac{\mu}{2}+\frac{1}{2})}}  F\left( \frac{\nu}{2}-\frac{\mu}{2}+1, -\frac{\mu}{2}-\frac{\nu}{2}+\frac{1}{2};\frac{3}{2};z^2\right) \nonumber \\
-\frac{\sin\frac{(\nu+\mu)\pi}{2} \Gamma{(\frac{\nu}{2}+\frac{\mu}{2}+\frac{1}{2}})}{2\Gamma{(\frac{\nu}{2}-\frac{\mu}{2}+1})} F\left( -\frac{\mu}{2}+\frac{\nu}{2}+\frac{1}{2}, -\frac{\mu}{2}-\frac{\nu}{2};\frac{1}{2};z^2\right)\bigg], \;\; -1<z <1. \quad
\end{eqnarray}
We find that it is also undefined at $z= \pm 1$ and so we consider the solution $Q_\nu^{-\mu}$, where it is given by
\begin{eqnarray}
Q_\nu^{-\mu}=2^{-\mu} \sqrt{\pi} (1-z^2)^{\frac{\mu}{2}}\bigg[\frac{\cos{\frac{(\nu-\mu) \pi}{2}} \Gamma{(\frac{\nu}{2}-\frac{\mu}{2}+1)}z}{\Gamma{(\frac{\nu}{2}+\frac{\mu}{2}+\frac{1}{2})}}  F\left( \frac{\nu}{2}+\frac{\mu}{2}+1, \frac{\mu}{2}-\frac{\nu}{2}+\frac{1}{2};\frac{3}{2};z^2\right) \nonumber \\
 -\frac{\sin\frac{(\nu-\mu)\pi}{2} \Gamma{(\frac{\nu}{2}-\frac{\mu}{2}+\frac{1}{2}})}{2\Gamma{(\frac{\nu}{2}+\frac{\mu}{2}+1})} F\left( \frac{\mu}{2}+\frac{\nu}{2}+\frac{1}{2},\frac{\mu}{2}-\frac{\nu}{2};\frac{1}{2};z^2\right)\bigg]
\quad -1<z <1,
\label{qfor}
\end{eqnarray}
Analyzing the above form, we find that the polynomial condition for $Q_\nu^{-\mu}$ is
\begin{eqnarray}
\nu=-(n+\mu+1), \quad n=0,1,2,3,...
\end{eqnarray}
Thus when $n$ is even, the first hypergeometric term in Eq. (\ref{qfor}) vanishes and the second term provides polynomial solutions and when $n$ is odd, the second term in Eq. (\ref{qfor}) vanishes and the first term provides polynomial solutions. Thus the polynomial solutions of $Q_{\nu}^{-\mu}$ can now be expressed as
\begin{eqnarray}
Q_{-(n+\mu+1)}^{-\mu}=\frac{2^{-\mu} \sqrt{\pi} \sin \left(\frac{(n+2 \mu+1) \pi}{2}\right) \Gamma{\left(-\frac{n}{2}-\mu\right)}}{2 \Gamma{\left(\frac{-n+1}{2}\right)}} (1-z^2)^{\frac{\mu}{2}} F\left(-\frac{n}{2}, \, \mu+\frac{n}{2}+\frac{1}{2};\,\frac{1}{2}; \,z^2\right), \quad \nonumber \\  n=0,2,4,..., \quad \label{ev_Q} \\
Q_{-(n+\mu+1)}^{-\mu}=\frac{2^{-\mu} \sqrt{\pi} \cos \left(\frac{(n+2 \mu+1) \pi}{2}\right)  \Gamma{(\frac{1}{2}-\frac{n}{2}-\mu)}}{\Gamma{(\frac{-n}{2})}} z (1-z^2)^{\frac{\mu}{2}} F\left(\frac{(1-n)}{2}, \frac{n}{2}+\mu+1; \frac{3}{2};z^2\right), \nonumber \\ n=1,3,5...\quad \label{od_Q}
\end{eqnarray}
Now comparing the expressions given in Eqs. (\ref{ev_Q}) and (\ref{od_Q}) with the ones given in Eqs. (\ref{even_n}) and (\ref{odd_n}), we find that $Q_{-(n+\mu+1)}^{-\mu}$ differs from $P_{n+\mu}^{-\mu}$ by just multiplicative constant coefficients only.  Thus $Q_{-(n+\mu+1)}^{-\mu}$ cannot be linearly independent of $P_{n+\mu}^{-\mu}$.  We then conclude that the admissible solutions $\phi_n(z)$ for (\ref{legeq}) can now be written as 
\begin{eqnarray}
\phi_n(z)= N_n P_{n+\mu}^{-\mu}(z), \quad n=0,1,2,3,...,
\end{eqnarray}
where $N_n$ are the normalization constants.
Reverting to the transformation given earlier, the eigenfunctions to Eq. (\ref{sim_mlo}) can be written as 
\begin{eqnarray}
\psi_n(x)=N_n (1-\lambda x^2)^{\frac{\bar{\gamma}-\bar{\alpha}}{2}} P_{n+\mu}^{-\mu}(\sqrt{\lambda} x), \quad n=0,1,2,3,...
\label{wavf}
\end{eqnarray}
\par It is now clear that the eigenfunctions corresponding to the Mathews-Lakshmanan oscillator for all orderings can be represented in terms of the associated Legendre polynomials with non-integral degree and order (that is, the degree of the polynomials $n+\mu$ and order of the polynomials $-\mu$ can in general take non-integral values).  The dependence of the eigenfunctions on the ordering arises from the presence of the parameters $\bar{\gamma}$, $\bar{\alpha}$ and $\mu$.  From Eq. (\ref{wavf}), we find the presence of the term $(1-\lambda x^2)^{\frac{\bar{\gamma}-\bar{\alpha}}{2}}$.  Such a presence indicates that the non-Hermitian orderings with $\bar{\gamma}<\bar{\alpha}$ cannot provide the eigensolution that is finite everywhere inside the interval $[-\frac{1}{\sqrt{\lambda}},\frac{1}{\sqrt{\lambda}}]$.  They always lead to singular solutions.  One can also find from \cite{ml} that the eigenfunctions obtained here match with the ones obtained by Mathews and Lakshmanan.  
With the choice of $\nu=n+\mu$, we find that the energy eigenvalues of the system become discretized as 
\begin{small}
\begin{eqnarray}
E_n = \left(n +\frac{1}{2}\,\right) \hbar\,\sqrt{k + \lambda^2 \hbar^2 (4 \overline{\alpha\,\gamma} + (\bar{\gamma} - \bar{\alpha})^2)} +  \frac{\lambda\,\hbar^2}{2} n(n+1) - \frac{\lambda \hbar^2}{2} (\alpha + \gamma), \;\, n=0,1,2,... \quad\label{eigg2}
\end{eqnarray}
\end{small}
\par The obtained eigenvalues also match with the results reported in \cite{ml, cari}.  Depending on the ordering, the values of $\bar{\alpha}$, $\bar{\gamma}$ and $\bar{\alpha \gamma}$ can be chosen and Eq. (\ref{eigg2}) shows that the eigenvalue spectrum changes accordingly with the considered ordering.  Note that the eigenvalues reduce to that of the harmonic oscillator for $\lambda=0$. 
\subsubsection{Orthonormality and relation to Jacobi polynomials:}
\par After deducing the eigenfunctions of the system, the next step is to check the orthonormality of the obtained solutions.  For this purpose, we note that the associated Legendre differential equation given in (\ref{legeq}) corresponds to a Sturm-Liouville problem so that it is guaranteed that the obtained polynomials are orthogonal.  But to prove the orthonormality of the above associated Legendre polynomials with non-integral degree and order in a rigorous sense, we here exploit its relation with one of the classical orthogonal polynomials, namely the Jacobi polynomials.  To show the above, we recall the identities of hypergeometric functions \cite{book_ab},
\begin{small}
\begin{eqnarray}
F(a,b;\frac{1}{2};z)&=&\frac{\Gamma(a+\frac{1}{2})\Gamma(b+\frac{1}{2})}{2 \sqrt{\pi} \Gamma({a+b+\frac{1}{2}})}\bigg[F\left(2a,2b;a+b+\frac{1}{2};\frac{1+\sqrt{z}}{2}\right) \nonumber \\&&+F\left(2a,2b;a+b+\frac{1}{2};\frac{1-\sqrt{z}}{2}\right)\bigg], \qquad \qquad \label{f12} \\
F(a,b;\frac{3}{2};z)&=&\frac{\Gamma(a-\frac{1}{2})\Gamma(b-\frac{1}{2})}{4 \sqrt{\pi} \Gamma({a+b-\frac{1}{2}})}\frac{1}{\sqrt{z}}\Bigg[F\left(2a-1,2b-1;a+b-\frac{1}{2};\frac{1+\sqrt{z}}{2}\right) \nonumber \\& & -F\left(2a-1,2b-1;a+b-\frac{1}{2};\frac{1-\sqrt{z}}{2}\right)\Bigg].\qquad \qquad \label{f12}
\end{eqnarray}
\end{small}
Using these identities in Eqs. (\ref{even_n}) and (\ref{odd_n}), we get
\begin{small}
\begin{eqnarray}
P_{n+\mu}^{-\mu}=\frac{2^{-(\mu+1)} (1-z^2)^{\frac{\mu}{2}}}{\Gamma{(\mu+1)}}\left[F\left(-n, (n+2 \mu+1); \mu+1; \frac{1+z}{2}\right)+F\left(-n, (n+2 \mu+1); \mu+1; \frac{1-z}{2}\right)\right], \nonumber \\  n=0,2,4,... \quad \label{pn_ev}\\
P_{n+\mu}^{-\mu}=\frac{-2^{-(\mu+1)} (1-z^2)^{\frac{\mu}{2}}}{\Gamma{(\mu+1)}}\left[F\left(-n, (n+2 \mu+1); \mu+1; \frac{1+z}{2}\right)-F\left(-n, (n+2 \mu+1); \mu+1; \frac{1-z}{2}\right)\right], \nonumber \\  n=1,3,5,... \quad \label{pn_od}
\end{eqnarray}
\end{small}
Note the above form of hypergeometric functions have relationship with the orthogonal Jacobi polynomials as
\begin{eqnarray}
F\left(-n,\,a+b+n+1,\,b+1,\, \frac{1+z}{2}\right)&=&\frac{\Gamma(n+1) \Gamma{(-b-n)}}{\Gamma(-b)} P_n^{(a,b)}(z), \\
F\left(-n,\,a+b+n+1,\,a+1,\, \frac{1-z}{2}\right)&=&\frac{\Gamma(n+1) \Gamma(a+1)}{\Gamma(a+n+1)} P_n^{(a,b)}(z),
\end{eqnarray}
where $P_n^{(a,b)}(z)$ represents the Jacobi polynomial. Using the above connections to Eqs. (\ref{pn_ev}) and (\ref{pn_od}), we find that the associated Legendre polynomials with non-integral degree and order ($P_{n+\mu}^\mu$) can be related to Jacobi polynomials ($P_n^{(a,b)}$) as
\begin{eqnarray}
P_{n+\mu}^{-\mu}(z)=\frac{ (-1)^n 2^{-\mu}\Gamma{(n+1)}}{\Gamma(n+\mu+1)}(1-z^2)^\frac{\mu}{2} P_n^{(\mu,\mu)}(z), \quad n=0,1,2,...
\label{pn_ja}
\end{eqnarray}
To deduce the above relation, one will have to use the simplification
\begin{eqnarray}
(-1)^n\frac{\Gamma(\mu+1)}{\Gamma(n+\mu+1)}+\frac{\Gamma{-(n+\mu)}}{\Gamma{(-\mu)}}=(-1)^n\frac{ 2\Gamma{(\mu+1)}}{\Gamma(n+\mu+1)}. 
\end{eqnarray}
\par The above mentioned relationship with the Jacobi polynomials (Eq. (\ref{pn_ja})) helps us to understand the properties of the associated Legendre polynomials with non-integral degree and order.  For example, one can now express the Rodrigues' formula of the latter from the former as
\begin{eqnarray}
P_{n+\mu}^{-\mu}= \frac{(-1)^n}{2^{n+\mu} (n+\mu)!} (1-z^2)^{\frac{-\mu}{2}}\frac{d^n}{dz^n}(1-z^2)^{n+\mu}.
\end{eqnarray}
\par Secondly, as the Jacobi polynomials are orthonormal in the interval $[-1,1]$, the obtained connection provides an easy way to find the orthonormality relation of the non-integral degree and order associated Legendre polynomials.  The orthonormality relation of the Jacobi polynomials is given by \cite{book_ab}
\begin{eqnarray}
\left( \frac{2^{a+b+1} \Gamma{(n+a+1)} \Gamma(n+b+1)}{(2n+a+b+1) \Gamma(n+a+b+1)n!} \right)^{-1} \int_{-1}^{1} (1-z)^a (1+z)^b P_n^{(a,b)}(z) P_m^{(a,b)}(z)dz=\delta_{nm}, \nonumber \\ \hspace{10cm} a,b>-1. \quad \; 
\end{eqnarray}
Using the above, we can deduce the orthonormality relation for the non-integral degree and order associated Legendre polynomials as 
\begin{eqnarray}
\left(\frac{2 n!  }{(2n+2\mu+1) (n+2\mu)!}\right)^{-1}\int_{-1}^{1}P_{n+\mu}^{-\mu}(z) P_{m+\mu}^{-\mu}(z) dz= \delta_{nm} , \quad \mu>-1
\label{ortho_pn}
\end{eqnarray}
\par Now, let us consider the normalization of the wavefunction given by Eq. (\ref{wavf}).  For this purpose, by rewriting the general ordered Schr\"odinger equation of the system (Eq. (\ref{sim_mlo})) as a Sturm-Liouville problem, we find that the weight function corresponding to the system is $(1-\lambda x^2)^{-({\bar{\gamma}-\bar{\alpha}})}$. Thus
\begin{small}
\begin{eqnarray}
\int_{-\frac{1}{\sqrt{\lambda}}}^{\frac{1}{\sqrt{\lambda}}} W(x) \psi_m^*(x) \psi_n(x) dx 
&=& N_n N_m\int_{-\frac{1}{\sqrt{\lambda}}}^{\frac{1}{\sqrt{\lambda}}} \ (1-\lambda x^2)^{-({\bar{\gamma}-\bar{\alpha}}) }  (1-\lambda x^2)^{{\bar{\gamma}-\bar{\alpha}}} P_{n+\mu}^{-\mu}(\sqrt{\lambda}x) P_{m+\mu}^{-\mu}(\sqrt{\lambda} x) dx\nonumber \\ 
&=&N_n N_m \frac{2 n! \delta_{nm} }{\sqrt{\lambda} (2n+2\mu+1) (n+2\mu)!}, \qquad \mu>-1
\end{eqnarray}
\end{small}
where the above orthonormalization puts forth a restriction that $\mu>-1$.  Note that here $\mu=\frac{1}{\hbar \lambda}\sqrt{k+\hbar^2 \lambda^2((\bar{\gamma}^2-\bar{\alpha}^2)+4 \overline{\alpha \gamma})}$.  The form of $\mu$ given above tells us that it is always positive as long as it is real, however there are orderings (or ordering parameters) that make $\mu$ to be imaginary.  For the latter cases, we cannot have bound state solutions.  By excluding the orderings corresponding to complex $\mu$, the normalized eigenstates for different orderings can be written as
\begin{eqnarray}
\psi_n(x)= \left(\frac{2 n! }{\sqrt{\lambda} (2n+2\mu+1) (n+2\mu)!}\right)^{\frac{-1}{2}}(1-\lambda x^2)^{\frac{(\bar{\gamma}-\bar{\alpha})}{2}} P_{n+\mu}^{-\mu}(\sqrt{\lambda}x) , \quad n=0,1,2,3... \quad
\end{eqnarray}
We again note that for the orderings $\bar{\gamma}<\bar{\alpha}$, the eigenfunctions become singular. 
\par Next, we represent the solution in terms of the $\lambda$-dependent Hermite polynomials which have been used in \cite{cari} to represent the eigenstates of this system for a particular non-Hermitian ordering. 
\subsubsection{$\lambda$-dependent Hermite polynomials}
\par As the considered Mathews-Lakshmanan  oscillator reduces to the harmonic oscillator in the limit $\lambda \rightarrow 0$, Carinena {\it et al} \cite{cari} have expressed the eigenstates of this system with modified Hermite polynomials.  For this purpose, they have reduced the Schr\"odinger equation in such a way that the  reduced equation becomes Hermite differential equation as $\lambda \rightarrow 0$.  Below, we show that the Schr\"odinger equation corresponding to all orderings can  be reduced to such a deformed Hermite differential equation and demonstrate that the eigenstates and spectrum obtained in this way is same as the one obtained in the previous subsection.  
\par For the above purpose, let us consider Eq. (\ref{sim_mlo}).  From the asymptotic results, we assume the solution to be of the form
\begin{eqnarray}
\psi(x )= (1 - \lambda x^2)^d \phi(x),\label{solt}
\end{eqnarray}
where
\begin{eqnarray}
d =\frac{ (\bar{\gamma} - \bar{\alpha})}{2} + \frac{\tilde{d}}{2\hbar \lambda}.
\end{eqnarray}
By doing so, we note that Eq. (\ref{sim_mlo}) gets reduced to the form
\begin{eqnarray}
\frac{d^2 \phi}{dx^2} - \frac{2(\frac{\tilde{d}}{\hbar}+\lambda)x}{(1 - \lambda x^2)} \frac{d \phi}{dx} + \frac{(b - \lambda(\bar{\gamma} - \bar{\alpha} )- \frac{\tilde{d}}{\hbar })}{(1 - \lambda x^2)} \phi(x) = 0.\label{par eq}
\end{eqnarray}
For the consideration of $\lambda$-dependent Hermite polynomials, the above equation has to be transformed in such a way that it will reduce to Hermite differential equation for $\lambda=0$. For this purpose, we rescale the independent variable and the parameter as,
\begin{eqnarray}
x = \sqrt{\frac{\hbar}{\tilde{d}}} y;  \quad
\lambda =  \frac{\tilde{d}}{\hbar} \tilde{\lambda}.
\end{eqnarray}
By doing so, we note that the Eq. $(\ref{par eq})$ gets reduced to the form
\begin{eqnarray}
\frac{d^2\phi}{dy^2} - \frac{ 2 (1 + \tilde{\lambda} ) y}{(1 - \tilde{\lambda} y^2)} \frac{d\phi}{dy} + \frac{B}{(1-\tilde{\lambda} y^2)}\phi = 0,\label{lam}
\end{eqnarray}
where
\begin{eqnarray}
B=\frac{\hbar}{\tilde{d}}\left(b-(\tilde{\lambda}(\bar{\gamma}-\bar{\alpha})+1)\frac{\tilde{d}}{\hbar}\right).
\end{eqnarray}
  The differential equation given in (\ref{lam}) is the $\lambda$-dependent Hermite differential equation. We here want to note that this $\lambda$-dependent Hermite differential equation is nothing but the Jacobi differential equation. To see this, we rewrite Eq. (\ref{lam}) with
\begin{eqnarray}
\sqrt{\tilde{\lambda}} y=z
\end{eqnarray}
so that it will become
\begin{eqnarray}
(1-z^2)\frac{d^2\phi}{dz^2} - \frac{2 (1 + \tilde{\lambda})}{\tilde{\lambda}} z \frac{d\phi}{dz} + \frac{B}{\tilde{\lambda}}\phi = 0
\label{lamm}
\end{eqnarray}
Now comparing this with the Jacobi differential equation,
\begin{eqnarray}
(1-z^2)\frac{d^2\phi}{dz^2}+(a-b-(a+b+2)z) \frac{d\phi}{dz}+n(n+a+b+1)\phi=0,
\label{jack}
\end{eqnarray}
we find that 
\begin{eqnarray}
a-b=0, \quad  a=\frac{1}{\tilde{\lambda}}=\mu, \quad \mathrm{and} \quad \frac{B}{\tilde{\lambda}}=n(n+2 \mu+1) 
\end{eqnarray}
Thus the solution of the $\lambda$-dependent Hermite differential equation can be written as
\begin{eqnarray}
\phi_n(y)=A_n{\cal{H}}_n(y; \tilde{\lambda})=A_n P_n^{(\mu, \mu)}(\sqrt{\tilde{\lambda}} y)
\end{eqnarray}
where $A_n$ are the normalization constants and $H_n(y, \tilde{\lambda})$ are the $\lambda-$ dependent Hermite polynomials. 
This relation shows that the energy eigenfunctions and eigenvalues obtained through these polynomials are the same as the ones obtained by associated Legendre polynomials with non-integral degree and order in the previous subsection. 
\subsection{Continuum states in region-$1$ and region-$3$ for $\lambda>0$}
\par As mentioned in Sec. 3 (wide Eqs. (\ref{boun1}) and (\ref{boun2})), one may not only have bound states in the region-2 ($-\frac{1}{\sqrt{\lambda}} \leq x \leq \frac{1}{\sqrt{\lambda}}$) as given by Eq. (\ref{boun1}), but there also exists the other possibility defined by (\ref{boun2}) which allows the solutions to vanish in the interior region (region-2) but they are non-zero in the outer regions (regions-1 and 3) (wide Eq. (\ref{boun2})). 
 For this purpose, one may first look for the existence bounded polynomial solutions in the regions-$1$ and $3$ and one can find that there exist no such solutions in these regions. 
\par For example, one can consider the form of $P_\nu^\mu$ defined in the region $1 \leq |z| \leq \infty$,
\begin{eqnarray}
P_\nu^\mu(z)=&\frac{2^{-(\nu+1)} \Gamma{(-\frac{1}{2}-\nu)}z^{-\nu+\mu-1}}{\sqrt{\pi}(z^2-1)^\frac{\mu}{2} \Gamma{(-\nu-\mu)}} F\left(\frac{1}{2}+\frac{\nu}{2}-\frac{\mu}{2},1+\frac{\nu}{2}-\frac{\mu}{2};\nu+\frac{3}{2};\frac{1}{z^2}\right) \nonumber \\
&+\frac{2^\nu \Gamma{(\frac{1}{2}+\nu)} z^{\nu+\mu}}{\sqrt{\pi}(z^2-1)^\frac{\mu}{2} \Gamma{(1+\nu-\mu)}} F\left(-\frac{\nu}{2}-\frac{\mu}{2}, \frac{1}{2}-\frac{\nu}{2}-\frac{\mu}{2};\frac{1}{2}-\nu;\frac{1}{z^2}\right).
\end{eqnarray}
As we did earlier, we here change $\mu$ by $-\mu$ so that the above function vanishes at $z = \pm 1$ and satisfies the boundary conditions at $z = \pm 1$.  The possible polynomial conditions for the above function are $\nu=n+\mu$ or $\nu=-n-\mu-1$.  As the associated Legendre differential equation is symmetric with respect to $\nu \rightarrow -(\nu+1)$, both of these polynomial conditions will not provide linearly independent polynomials.  Using the polynomial condition $\nu=n+\mu$, one can find that 
\begin{eqnarray}
P_{n+\mu}^{-\mu}(z)= \frac{2^{n+\mu} \pi^{-\frac{1}{2}} \Gamma{(n+\mu+1/2) (z^2-1)^{\frac{\mu}{2}}}}{\Gamma{(1+n+2 \mu)}} \sum_{i=0}^{n} a_i z^{i},
\label{pnmu_r}
\end{eqnarray}
where for odd values of $n$, $a_j=0$ for only $j=0,2,4...$ and $a_j\neq 0$ for $j=1,3,5...$.  For even values of $n$, $a_j=0$ for only $j=1,3,5,...$ and $a_j \neq 0$ for $j=0,2,4,...$. However the polynomial solution in Eq. (\ref{pnmu_r}) goes to infinity at $z= \pm \infty$ and so it is not finite as $z= \pm \infty$. Thus one cannot generate bounded polynomial solutions from $P_{n+\mu}^{-\mu}(z)$ in the regions $1 \leq |z| \leq \infty$.  Similarly, while considering the function $Q_\nu^\mu(z)$ which is of the form
\begin{small}
 \begin{eqnarray}
Q_\nu^{\mu}(z)=e^{i \mu \pi} 2^{-\nu-1} \pi^{\frac{1}{2}}{\frac{\Gamma{(\nu+\mu+1)}}{\Gamma(\nu+\frac{3}{2})}} z^{-\nu-\mu-1} (z^2-1)^\frac{\mu}{2} F\left(1+\frac{\nu}{2}+\frac{\mu}{2}, \frac{1}{2}+\frac{\nu}{2}+\frac{\mu}{2}; \nu+\frac{3}{2}; \frac{1}{z^2}\right), \quad
\label{qnumu_r}
\end{eqnarray}
\end{small}
the polynomial condition for the hypergeometric function present in the above equation is $\nu+\mu+1=-n$. But for this choice, the gamma function term in the numerator $\Gamma{(\nu+\mu+1)}$ becomes infinite.  Thus there are no bounded polynomial solutions in the regions-$1$ and $3$ mentioned in Sec. 3.
\par However in Refs. \cite{ml} and  \cite{joos}, it has been shown that for $\nu=\frac{-1}{2}+i \rho$, (where $\rho$ takes any arbitrary real value), the associated Legendre function satisfies the boundary condition given by (\ref{boun2}), that is 
\begin{eqnarray}
P_{-\frac{1}{2}+i \rho}^{-\mu}(z) &=0,  \quad -1\leq z \leq 1, \nonumber \\
& \neq  0, \quad 1< |z|< \infty. 
\end{eqnarray}
Thus one can represent the eigenfunctions in terms of these functions. 
Importantly, the above associated Legendre functions are found to satisfy \cite{joos,ml}
\begin{eqnarray}
\int_1^{\infty} P_{-\frac{1}{2}+i \rho}^{-\mu}(z) P_{-\frac{1}{2}+i \rho'}^{-\mu}(z)dz=\frac{(i\rho-1)!(-i \rho-1)!}{(i \rho+\mu-\frac{1}{2})!(-i \rho+\mu-\frac{1}{2})!}\delta(\rho-\rho'),
\label{intor}
\end{eqnarray}
where $\delta$ represents the Dirac delta function.  Thus these functions show interesting orthonormality relations for different values of $\rho$ and so one can represent orthonormal basis of eigenfunctions for this case as \cite{ml}
\begin{eqnarray}
\psi(x)=\left[\frac{(i\rho-1)!(-i \rho-1)!}{(i \rho+\mu-\frac{1}{2})!(-i \rho+\mu-\frac{1}{2})!}\right]^{-\frac{1}{2}} P_{-\frac{1}{2}+i\rho}^{-\mu}(\pm \sqrt{\lambda} x) \theta(\pm \sqrt{\lambda}x-1),
\label{conteig}
\end{eqnarray}
where $\theta$ is the unit step function: $\theta(x)=0$ for $x <0$, $\theta(x)=1$ for $x>0$. The positive sign in $P_{-\frac{1}{2}+i\rho}^{-\mu}(\pm \sqrt{\lambda} x)$ and in $\theta$ function denotes the localization in region 3 and the negative sign in them denotes the localization in region 1.  The energy eigenvalues of the system corresponding to this case are 
\begin{eqnarray}
E=\frac{-\hbar^2}{2}\left[\rho^2+\frac{1}{4}+(\bar{\alpha}+\bar{\gamma})+\mu^2\right], \quad -\infty < \rho < \infty.
\label{conten}
\end{eqnarray}
\par It may be noted that in the above even though $\nu$ $(=-\frac{1}{2}+i\rho)$ is complex, the energy eigenvalues are real.  Note that as the eigenfunctions given in Eq. (\ref{conteig}) are orthonormal to each other and that they vanish at $z \rightarrow \pm \infty$, these functions represent bound states.  Interestingly the energy eigenvalues given in Eq. (\ref{conten}) are found to be continuous even though the states are bounded.   Thus these states correspond to continuum bound states similar to the case of potentials obtained by amplitude modulating the free particle wavefunction  \cite{prapap}.  

\section{\label{laml0}Case: $\lambda<0$}
\par While considering the case $\lambda<0$, the system in Eq. (\ref{sim_mlo}) becomes
\begin{eqnarray}
(1+|\lambda|x^2) \frac{d^2 \psi}{dx^2} -2 a |\lambda|x \frac{d \psi}{dx}+\left(b+\frac{c x^2}{(1+|\lambda|x^2)}\right)\psi=0,
\label{eqn2}
\end{eqnarray}
where 
\begin{eqnarray}
a=\bar{\gamma}-\bar{\alpha}-1; \;\; b=-2|\lambda| \bar{\gamma}+\frac{2 E}{\hbar^2}; \;\; c=-(4 \overline{\alpha \gamma} \lambda^2+\frac{k}{\hbar^2}).
\end{eqnarray}
From (\ref{eqn2}), one can find that the system has no singularities other than $\pm \infty$ so that we look for the eigenstates which remain finite in the region $-\infty < x < \infty$ and vanish sufficiently fast as $x \rightarrow \pm \infty$. Through the transformations 
\begin{eqnarray}
\psi=(1-z^2)^{(\bar{\gamma}-\bar{\alpha})/2} \phi(z), \quad z=i\sqrt{|\lambda|} x,
\label{tr_din}
\end{eqnarray}
one can reduce Eq. (\ref{eqn2}) to the associated Legendre differential equation (\ref{legeq}), where $\nu(\nu+1)$ and $\mu$ present therein can be represented as 
\begin{eqnarray}
\nu(\nu+1)=-\frac{b}{|\lambda|}-(\bar{\gamma}-\bar{\alpha})+\mu^2,\quad \mu=\frac{\tilde{d}}{\hbar |\lambda|}, \quad
\tilde{d}=\sqrt{k+\hbar^2 \lambda^2(4 \overline{\alpha \gamma}+(\bar{\gamma}-\bar{\alpha})^2)}.
\label{cons2}
\end{eqnarray}
Let $z=iy$ so that the eigenstates of the system can be represented either by $P_\nu^\mu(iy)$ or by $Q_\nu^\mu(iy)$.  

 \par In this connection, the functions $P_\nu^\mu(iy)$ and $Q_\nu^\mu(iy)$ can be expressed as \cite{book_ab} 
\begin{small}
\begin{eqnarray}
P_\nu^\mu(iy)=\frac{2^{-(\nu+1)}\, \Gamma{(-\nu-\frac{1}{2})}\, (-y^2)^{\frac{-\nu+\mu-1}{2}}}{\sqrt{\pi}\, (-(1+y^2))^\frac{\mu}{2}\, \Gamma{(-\nu-\mu)}} F\left(\frac{1}{2}+\frac{\nu}{2}-\frac{\mu}{2},\, 1+\frac{\nu}{2}-\frac{\mu}{2}; \,\nu+\frac{3}{2}; \, \frac{-1}{y^2}\right) \qquad \qquad \qquad \quad \nonumber \\
+\frac{2^\nu\, \Gamma{(\nu+\frac{1}{2})}\, (-y^2)^{\frac{\nu+\mu}{2}}}{\sqrt{\pi}\,\Gamma(1+\nu-\mu)\, (-(1+y^2))^{\frac{\mu}{2}}} F\left(\frac{-(\nu+\mu)}{2},\, \frac{1}{2}-\frac{\nu}{2}-\frac{\mu}{2};\, \frac{1}{2}-\nu ;\, \frac{-1}{y^2}\right), \qquad \label{p1} \\
 Q_{\nu}^{\mu}(iy)=\frac{e^{i \mu \pi}\, 2^{-(\nu+1)}\, \sqrt{\pi}\, \Gamma{(1+\nu+\mu)} \,(-(1+y^2))^{\frac{\mu}{2}}}{\Gamma{(\nu+\frac{3}{2})}\,(-y^2)^{\frac{\nu+\mu+1}{2}}}  F\left(\frac{1}{2}+\frac{\nu}{2}+\frac{\mu}{2},\,1+\frac{\nu}{2}+\frac{\mu}{2}; \, \nu+\frac{3}{2}; \, \frac{-1}{y^2}\right). \qquad \label{q1}
\end{eqnarray}
\end{small}
Note that the above forms of $P_\nu^\mu(iy)$ and $Q_\nu^\mu(iy)$ are singular at $y=0$.  But here we require a solution which is finite over $-\infty \leq y \leq \infty$ so that we carry out a Pfaff transformation as \cite{book_ab}
\begin{eqnarray}
F\left(a,b;c;z\right)=(1-z)^{-a}\, F\left(a,\, c-b;\, c;\, \frac{z}{z-1}\right).
\end{eqnarray}
By doing the above transformation in Eqs. (\ref{p1}) and (\ref{q1}), we will get $P_\nu^\mu(iy)$ and $Q_\nu^\mu(iy)$ as
\begin{small}
\begin{eqnarray}
P_\nu^\mu(iy)&=&\frac{2^{-(\nu+1)} \, \Gamma{(-\nu-1/2)} \, (-(1+y^2))^{-(\nu+1)/2}}{\sqrt{\pi} \,\Gamma{(-\nu-\mu)}}
                F\left(\frac{1}{2}+\frac{\nu}{2}-\frac{\mu}{2},\, \frac{1}{2}+\frac{\nu}{2}+\frac{\mu}{2}; \, \nu+\frac{3}{2}; \, \frac{1}{1+y^2}\right) \nonumber \\
              &+& \frac{2^\nu \, \Gamma{(\nu+1/2)} \, (-(1+y^2))^{\frac{\nu}{2}}}{ \sqrt{\pi}\,\Gamma{(1+\nu-\mu)}}
                F\left(  -\left(\frac{\nu}{2}+\frac{\mu}{2}\right), \, -\frac{\nu}{2}+\frac{\mu}{2}; \, \frac{1}{2}-\nu; \, \frac{1}{1+y^2}\right) \quad
\label{pj}
\end{eqnarray}
\end{small}
\begin{eqnarray}
Q_\nu^\mu(iy)=&\frac{e^{i \mu \pi} 2^{-(1+\nu)}\, \sqrt{\pi}\,  \Gamma{(1+\nu+\mu)}\, (-(1+y^2))^{\frac{-(1+\nu)}{2}}}{\Gamma{(\nu+3/2)}} F\left(\frac{1}{2}+\frac{\nu}{2}+\frac{\mu}{2},\frac{1}{2}+\frac{\nu}{2}-\frac{\mu}{2}; \nu+\frac{3}{2}; \frac{1}{(1+y^2)} \right) \qquad
\label{qj}
\end{eqnarray}
In the above, we expanded the hypergeometric series in terms of $\frac{1}{(1+y^2)}$ so that the solution remains finite everywhere in the region $|y| \leq \infty$ and these hypergeometric functions tend to zero as $y \rightarrow \pm \infty$. We again look for bounded polynomial solutions to the above Eqs. (\ref{pj}) and (\ref{qj}). 
\par For the purpose, the above $P_\nu^\mu(iy)$ can be related to orthonormal Gegenbauer polynomials for the choices
\begin{eqnarray}
\nu=n-\mu, \quad \mathrm{or} \quad \nu=-n+\mu-1,
\label{pold}
\end{eqnarray}
with the associated relation between the hypergeometric function and Gegenbauer polynomial \cite{book_ab}
\begin{eqnarray}
F\left(-\frac{n}{2},b; b+\frac{1-n}{2};z\right)=\frac{(-1)^n \, n!}{(1-2b)_n}\, \, C_n^{b-\frac{n}{2}}(\sqrt{1-z}).
\end{eqnarray}
Thus, $P_{n-\mu}^{\mu}$ and $P_{-n+\mu+1}^{\mu}$ can give rise to orthonormal sets of solutions.  However, as mentioned in Eq. (\ref{gg}), due to the symmetry of associated Legendre differential equations with respect to $\nu \rightarrow -(\nu+1)$, $P_{n-\mu}^{\mu}$ and $P_{-n+\mu+1}^{\mu}$ cannot be linearly independent solutions.  Even if we look for bounded polynomial solutions from $Q_\nu^\mu$, one cannot find any additional linearly independent solution. So we consider only $P_{n-\mu}^{\mu}$ and it can be related to the Gegenbauer polynomials as
\begin{eqnarray}
P_{n-\mu}^{\mu}(iy)=\frac{(-1)^{\frac{n-\mu}{2}}\, 2^{n-\mu} \, n!\, \Gamma{(n-\mu+\frac{1}{2})}}{\sqrt{\pi}\,\Gamma{(1+2n-2\mu)}} (1+y^2)^{\frac{n-\mu}{2}}C_n^{\mu-n} \left(\frac{y}{\sqrt{1+y^2}}\right).
\label{gegen}
\end{eqnarray}
Thus the eigenfunctions of the system for this case can be written as
\begin{eqnarray}
\psi_n(x) =N_n (1+|\lambda|x^2)^{\frac{\bar{\gamma}-\bar{\alpha}}{2}} P_{n-\mu}^\mu(i\sqrt{|\lambda|}x),
\label{pnn}
\end{eqnarray}
where, $N_n$ represents the normalization constant. The energy eigenvalues of the system for this case are
\begin{eqnarray}
E_n=-n(n+1)\frac{\hbar^2 |\lambda|}{2}+(n+\frac{1}{2})\hbar \tilde{d}+\frac{\hbar^2 |\lambda|}{2}(\bar{\gamma}+\bar{\alpha}).
\end{eqnarray} 
The eigenfunctions are found to be orthonormal with respect to the weight function $W(x)=(1+|\lambda|x^2)^{-\frac{\bar{\gamma}-\bar{\alpha}}{2}}$.  One can find that for $m \neq n$, 
\begin{small}
\begin{eqnarray}
\int_{-\infty}^{\infty} P_{n-\mu}^\mu(i\sqrt{|\lambda|x}) P_{m-\mu}^{\mu} (i\sqrt{|\lambda|x}) dx&=& \int_{-\infty}^{\infty} (1+|\lambda|x^2)^{\frac{n+m}{2}-\mu} C_n^{\mu-n} \left(\frac{\sqrt{|\lambda|}x}{\sqrt{1+|\lambda| x^2}}\right)C_m^{\mu-m} \left(\frac{\sqrt{|\lambda|}x}{\sqrt{1+|\lambda| x^2}}\right) dx \nonumber \\
&=&\, 0,  \qquad \qquad  \qquad \qquad \mu > \frac{m+n+1}{2},
\end{eqnarray}
\end{small}
and for $m=n$,
\begin{small}
\begin{eqnarray}
\int_{-\infty}^{\infty} P_{n-\mu}^\mu(i\sqrt{|\lambda|}x) P_{n-\mu}^{\mu} (i\sqrt{|\lambda|}x) dx=\frac{4 \, n!\, \sin ^2((n-\mu)\pi ) \Gamma (2 \mu -n)}{ (2n+1-2 \mu)\pi}, \quad \mu> n+\frac{1}{2}.
\label{or_pnn}
\end{eqnarray}
\end{small}
Thus the normalization constant $N_n$ in (\ref{pnn}) takes the form
\begin{eqnarray}
N_n=\left(\frac{4 \, n!\, \sin ^2((n-\mu)\pi ) \Gamma (2 \mu -n)}{ (2n+1-2 \mu)\pi}\right)^{-\frac{1}{2}}.
\end{eqnarray}
The restriction $\mu>n+\frac{1}{2}$ in Eq. (\ref{or_pnn}) indicates the energy eigenspectrum corresponding to this case is finite and that $n$ can take integral values only between $0$ and $N$, where $N$ is the maximal integer less than $\mu-\frac{1}{2}$.  

\section{\label{sumry} Summary}
\par In this paper, we have studied the quantum exact solvability of the general ordered position dependent mass problem of the one-dimensional Mathews-Lakshmanan oscillator.  We have represented the eigenfunctions of the system in terms of the interesting non-integral degree and order associated Legendre polynomials and explored their characteristics.  Such associated Legendre polynomials have been found to form orthonormal basis of functions.  We have also shown the inter-relation between such non-integral degree and order associated Legendre polynomials with the $\lambda$-dependent Hermite polynomials and Jacobi polynomials.   We have also studied the quantum solvability of the considered problem for two different situations, namely $\lambda>0$ and $\lambda<0$.  Further, we have discussed an interesting situation in which the Mathews-Lakshmanan oscillator is found to support bound states with continuous energy values.  For all these problems the associated energy spectra were also deduced. 
\section*{Acknowledgements}
SK thanks the Department of Science and Technology (DST), Government of India, for providing a INSPIRE Fellowship.   The work of MS forms part of a research project sponsored by Department of Science and Technology, Government of India.  The work of ML is supported by a NASI Senior Scientist Platinum Jubilee fellowship program.  He is also grateful to the Science and Engineering Research Board (SERB) for support through a research project.

\end{document}